\documentclass[aps,prl,twocolumn, reprint,superscriptaddress]{revtex4}
\usepackage{graphicx}

\begin{document}

\title{Machine Learning Enabled Lineshape Analysis in Optical Two-Dimensional Coherent Spectroscopy}


\author{Srikanth Namuduri}
\affiliation
{Department of Electrical and Computer Engineering, Florida International University, Miami, Florida 33199, USA}

\author{Michael Titze}
\affiliation
{Department of Physics, Florida International University, Miami, Florida 33199, USA}
\affiliation
{Sandia National Laboratories, Albuquerque, New Mexico 87185, USA}

\author{Shekhar Bhansali}
\affiliation
{Department of Electrical and Computer Engineering, Florida International University, Miami, Florida 33199, USA}

\author{Hebin Li}
\email[Email address: ]{hebin.li@fiu.edu}
\affiliation
{Department of Physics, Florida International University, Miami, Florida 33199, USA}

\begin{abstract}
Optical two-dimensional (2D) coherent spectroscopy excels in studying coupling and dynamics in complex systems. The dynamical information can be learned from lineshape analysis to extract the corresponding linewidth. However, it is usually challenging to fit a 2D spectrum, especially when the homogeneous and inhomogeneous linewidths are comparable. We implemented a machine learning algorithm to analyze 2D spectra to retrieve homogeneous and inhomogeneous linewidths. The algorithm was trained using simulated 2D spectra with known linewidth values. The trained algorithm can analyze both simulated (not used in training) and experimental spectra to extract the homogeneous and inhomogeneous linewidths. This approach can be potentially applied to 2D spectra with more sophisticated spectral features.  
\end{abstract}

\maketitle
Optical two-dimensional coherent spectroscopy (2DCS) \cite{Cundiff2013,Li2017} has become a powerful technique to study coupling and dynamics in complex quantum systems such as atomic ensembles \cite{Tian2003,Tekavec2007,Dai2012,Li2013,Gao:16,PhysRevLett.120.233401,Yu2019,Yu2018}, semiconductor quantum wells \cite{Li2006a,Stone2009a,Cundiff2012,Turner2012,Singh2013,PhysRevLett.112.097401,Nardin2014} and dots \cite{PhysRevB.87.041304,Moody2013,Moody2013a,Moody2013b}, 2D materials \cite{Moody2015,Titze2018}, perovskites \cite{Monahan2017,Richter2017,Jha2018,Thouin2018,Nishida2018,Titze2019}, and photosynthesis \cite{Brixner2005,Engel2007,Collini2010}. By unfolding a spectrum onto a 2D plane, 2DCS can better isolate and probe the dynamics of different processes that might be difficult to separate otherwise. After the Fourier transform, the temporal dynamics are transferred into the spectral lineshape in the frequency domain. Thus it is critical to establish an effective lineshape analysis of 2D spectra to extract accurate linewidths including both homogeneous and inhomogeneous linewidths. An analytical solution based on the projection-slice theorem has been developed \cite{Siemens2010} to fit a 1D slice in the diagonal and/or cross-diagonal direction. This method only analyzes 1D slices but does not fully utilize the entire 2D spectrum and stronger constraint associated with it. In an improved approach \cite{Bell2015}, an analytical form of 2D spectra can be used to perform a 2D fit of the whole spectrum. The 2D fit might be difficult to converge when there is a noise background. In this letter, we propose an alternative lineshape analysis approach to analyze 2D spectra by using a machine learning algorithm. 

Machine learning has been a fast growing field in the past decade and made remarkable progress in well known applications such as image processing, speech recognition, and self-driving cars, among others. Meanwhile, machine learning also has a profound impact on solving scientific problems. For instance, machine learning has been implemented to solve quantum many-body problems \cite{Carleo2017}, perform quantum state tomography \cite{Torlai2018}, and search for new materials with targeted properties \cite{Xue2016}. In optical spectroscopy, machine learning has been used to analyze spectra for classification and to predict spectra based on the molecular structure \cite{Ghosh2019}. 

In this work, we implemented a deep learning algorithm based on a convolutional neural network to extract homogeneous and inhomogeneous linewidths from 2D spectra. The network was first trained with a group of simulated 2D spectra and the corresponding linewidth values. The trained algorithm was then used to analyze a different set of simulated 2D spectra to evaluate its effectiveness in extracting the linewidths. Finally, we implemented the algorithm on two experimental 2D spectra to measure their homogeneous and inhomogeneous linewidths and compare them with values obtained from the analytical fit of slices.

\begin{figure}[htbp]
\centering
\includegraphics[width=\linewidth]{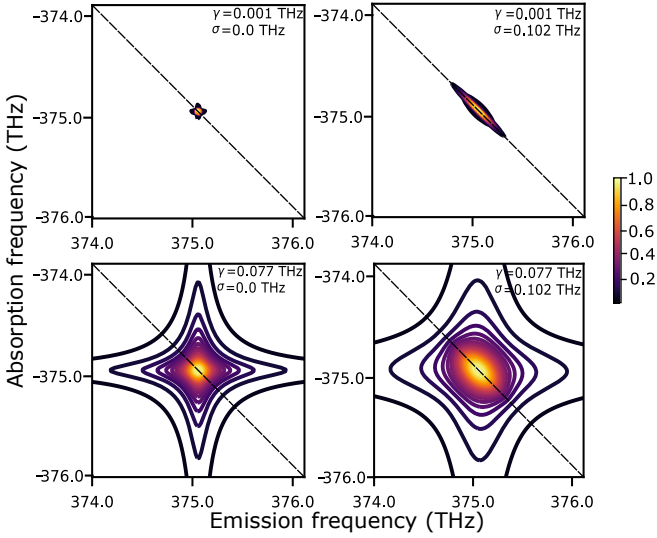}
\caption{Examples of simulated 2D spectra of a two-level system with different values of homogeneous linewidth $\gamma$ and inhomogeneous linewidth $\sigma$.}
\label{fig:samples}
\end{figure}

We consider a single 2D spectral resonance resulted from the electronic transition of a two-level system under the excitation of three pulses  \cite{Li2017}. The three pulses with wave vectors $\mathbf{k}_A$, $\mathbf{k}_B$, and $\mathbf{k}_C$ arrive at the sample in that time order. The time delay between the first and second pulses is $\tau$, the time delay between the second and third pulses is $T$, and the signal emission time is $t$. By solving the optical Bloch equations to the third order with the rotating wave approximation, delta-function pulses and zero waiting time $T=0$, we obtain the transient Four-wave mixing signal in the phase-matching direction $\mathbf{k}_S=-\mathbf{k}_A+\mathbf{k}_B+\mathbf{k}_C$ as a function of $\tau$ and $t$ \cite{Yajima1979}
\begin{equation}
s(\tau, t) = s_0 e^{-(\gamma+i \omega_0)\tau-(\gamma-i\omega_0)t+\sigma^2(t-\tau)^2/2}\Theta(\tau)\Theta(t), \label{eq1}
\end{equation}
where $s_0$ is the amplitude at time zero, $\omega_0$ is the resonance frequency, $\gamma$ is the homogeneous linewidth corresponding to the dephasing rate, $\sigma$ is the inhomogeneous linewidth, and the $\Theta$'s are Heaviside step functions. For a given combination of $\gamma$ and $\sigma$, the time-domain signal $s(\tau,t)$ is calculated according to Eq. (\ref{eq1}) and subsequently Fourier transformed into the frequency domain to generate a 2D spectrum $s(\omega_\tau, \omega_t)$. Some typical 2D spectra are shown in Fig. \ref{fig:samples} for different values of $\gamma$ and $\sigma$. A set of 4096 spectra is generated with 64 values of $\gamma$ and $\sigma$ each with equal spacing in the range $\gamma= 0.001 \sim 0.3$ THz and $\sigma= 0.0\sim0.2$ THz. The generated spectra are randomly divided into two groups with $90\%$ in the training group and $10\%$ in the testing group. 

The training group of simulated 2D spectra with known $\gamma$ and $\sigma$ values is used to train a neural network algorithm to extract the homogeneous and inhomogeneous linewidths from a 2D spectrum. During the training process, the neural network algorithm learns a functional relationship between the shape of the simulated 2D spectra and the corresponding linewidths. For this purpose, the network, which consists of processing units known as nodes, is organized as layers. Each layer learns an intermediate computation that is required to represent the overall functional relationship. The layered arrangement of the nodes enables the network to learn a complex function as a sequence of simpler functions. The layers in deep learning neural networks also provide multiple levels of representation of the input data, with each layer transforming the output of the previous layer into a more abstract representation \cite{Goodfellow-et-al-2016, lecun2015deep}. The algorithm can also include convolutional layers to implement a convolutional neural network (CNN) \cite{Goodfellow-et-al-2016, lecun2015deep} which is commonly used in analyzing high-resolution images with a large number of pixels to reduce the computational requirement. Although the spectra analyzed here only have a modest number of pixels ($128\times 128$), implementing a CNN is important for analyzing high-resolution 2D spectra in general. 

\begin{figure}[htbp]
\centering
\includegraphics[width=\linewidth]{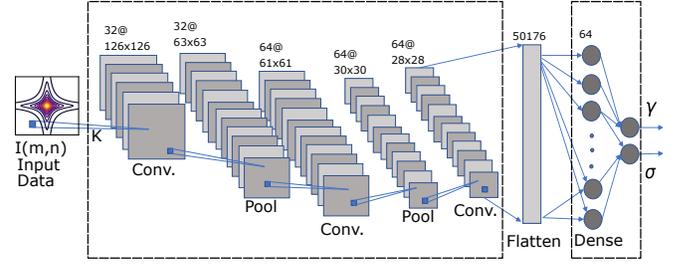}
\caption{Schematic of the neural network architecture. The network includes convolutional (Conv.) layers, pooling (Pool) layers, a flatten layer, a dense layer, and the output layer.}
\label{fig:network}
\end{figure}

The architecture of our learning algorithm is summarized in the schematic shown in Fig. \ref{fig:network}. The input data for each training is a simulated 2D spectrum represented by a 2D array $I(m,n)$ of size $128\times 128$. The neural network consists of two main stages as outlined by the dotted lines in Fig. \ref{fig:network}. The first stage is composed of three convolutional layers interspersed with two pooling layers. The output of the final convolutional layer is converted into a 1D array, referred to as the flattened layer, and subsequently input to the dense layer in the second stage. In the dense layer, the outputs of the nodes are calculated from the flattened layer by computing a weighted sum of all the inputs and applying an activation function. A similar calculation is done with the dense layer to generate the inputs for two nodes in the output layer corresponding to the extracted homogeneous and inhomogeneous linewidths.  

In a convolutional layer, the input array is convoluted with a set of learnable filters by computing the dot product between the filter and the input to produce a 2D activation map of that filter. If the input array is represented by $I$ and the filter by $K$, then the output of the convolution operation $C$ is given by 

\begin{equation}
    C(i,j) = \sum_m \sum_n I(m,n)K(i-m,j-n)
    \label{eq2}
\end{equation}

The output of a convolutional layer, referred to as a feature map, is a stack of the activation maps of all filters. When a specific feature associated with a spatial position in the input array is detected, the network learns the filters that activate such a feature. The output of a convolutional layer can be processed by pooling, which is effectively a form of nonlinear down-sampling to reduce the dimensions of the data. A pooling layer partitions the input array into small rectangular regions ($2\times2$ in our case) and maps their aggregate value to a single pixel in the next layer. We use the max pooling which outputs the maximum value of each region. This step also makes the network more robust to the noise in the data, besides reducing the data dimensions. As shown in Fig. \ref{fig:network}, our implementation consists of three convolutional layers using 32 filters for the first and 64 filters for the others. The first two convolutional layers each are followed by a pooling layer which reduces both dimensions of the input array by half. The output of the third convolutional layer is flattened into a 1D array as an input for the dense layer. 

The dense layer uses a fully connected network. In general, if the number of layers is $M$, the number of nodes in the $m^{th}$ layer is $N_h^m$ and the weight and bias vectors of the $m^{th}$ layer are given by $W_{i,j}^{m}$ and $b_j^m$, respectively, then the output of the $m^{th}$ layer is obtained by the two following equations:

\begin{equation}
    h_i^m = \sum_{j=1}^{N_h^{m-1}}  W_{i,j}^{m}.y_j^{m-1} + b_j^m,
    \label{eq3}
\end{equation}

\begin{equation}
    y_i^m = A(h_i^m).
    \label{eq4}
\end{equation}
In Eq. (\ref{eq4}), the function $A$ represents the activation function, which in this case is the rectified linear unit (ReLU) given by $A(z) = max(z,0)$. The output of the overall network is obtained by calculating the outputs from each layer using the above equations. We note that the final output layer includes two nodes that give the extracted values of the homogeneous and inhomogeneous linewdiths of a 2D spectrum.  

The process of training a neural network algorithm involves optimizing the weights (filters for the convolutional layers) in each layer to minimize the loss function. The loss function is an estimate of the overall error in the functional relationship learned by the network. The error for each input training spectrum is the difference between the extracted and actual values of the linewidths. The extracted values are calculated using the network illustrated in Fig. \ref{fig:network} \cite{rodriguez2019machine, 8788580}. If $y_{extr}$ is the value of the extracted linewidth and $y_{actual}$ is the actual value, then the loss function used in this research is given by
\begin{equation}
    L = \frac{1}{N} \sum_1^N (y_{extr} - y_{actual})^2
    \label{eq6}
\end{equation}
Optimizing the weights of a neural network is mathematically equivalent to finding the global minima. For this purpose, the partial derivatives of the loss function are calculated with respect to the input variables and are used to update the respective weights. The partial derivatives or gradients are computed using the backpropagation algorithm \cite{lecun1989backpropagation, bottou2010large}.

In this work, the neural network algorithm described in Fig. \ref{fig:network} was implemented in Python by using the Keras library. The computation was carried out on Google Colaboratory, which is a free computational environment, with access to a Tesla K80 GPU from Nvidia with 12GB of RAM. A total number of 3686 spectra were randomly selected from 4096 simulated spectra as the training data. The neural network was trained for 200 epochs with a learning rate of $10^{-5}$. 

\begin{figure}[htbp]
\centering
\includegraphics[width=0.9\linewidth]{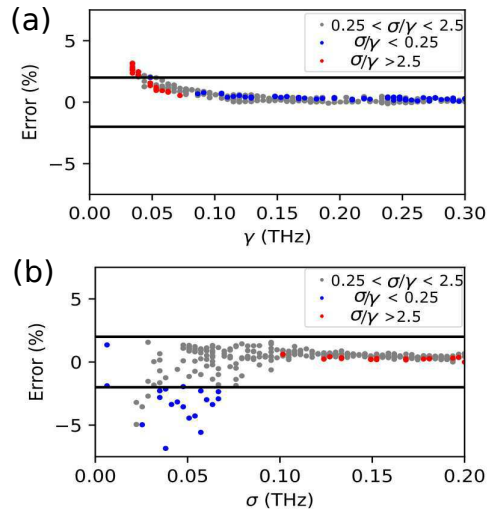}
\caption{Percentage error in the (a) homogeneous and (b) inhomogeneous linewidth values extracted from the 410 simulated 2D spectra that were not used for training. The data points are color coded for different ranges of ratio $\sigma/\gamma$.}
\label{fig:errors}
\end{figure}

After training the network, we first tested the algorithm with a set of 410 simulated spectra that were not used for the training. Each testing spectrum was analyzed by the trained algorithm to extract the values of both $\gamma$ and $\sigma$ which were compared to the actual values of the linewidths to calculate the $R^2$ value and the percentage error. The obtained $R^2$ value is 0.995. The percentage errors are plotted in Fig. \ref{fig:errors} for different values of both (a) the homogeneous linewidth $\gamma$ and (b) the inhomogeneous linewidth $\sigma$. Most of the data points are within the percentage error of $\pm2\%$ or smaller. In general, the percentage error increases as the linewidth value decreases since the denominator becomes smaller and the spectra have a resolution limited by the number of pixels. In addition, the ratio of the inhomogeneous and homogeneous linewidths, $\sigma / \gamma$, strongly affects the percentage error. The data points in Fig. \ref{fig:errors} are color coded for different values of $\sigma / \gamma$: blue for $\sigma/\gamma<0.25$, red for $\sigma / \gamma>2.5$, and gray for $0.25 \leq \sigma / \gamma \leq 2.5$. The inhomogeneous linewidth has greater errors, up to $5\%$, for the blue data points corresponding to the spectra that are strongly homogeneous, i.e. $\gamma$ is much larger than $\sigma$. On the other hand, for the red data points representing the spectra that are strongly inhomogeneous, i.e. $\sigma$ is much larger than $\gamma$, the homogeneous linewidth has relatively larger percentage errors. However, the algorithm performs well for the spectra with comparable $\gamma$ and $\sigma$, the intermediate cases that are difficult to analyze by fitting \cite{Siemens2010}. 

\begin{figure}[bt]
\centering
\includegraphics[width=0.9\columnwidth]{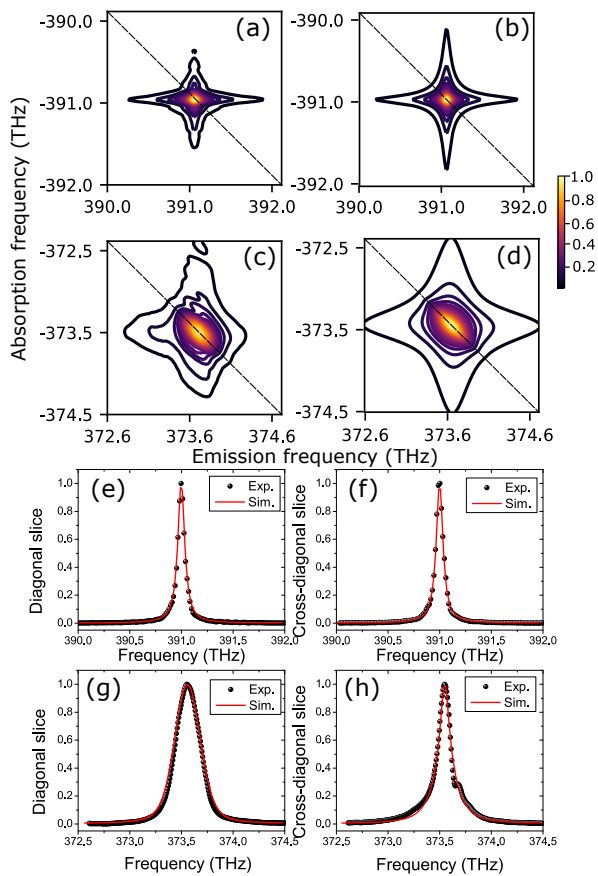}
\caption{(a) Experimental and (b) simulated 2D spectra of a K vapor. (c) Experimental and (d) simulated 2D spectra of a quantum well. Simulated 2D spectra are generated with the linewidths extracted from experimental spectra. (e,g) Diagonal slices and (f, h) cross-diagonal slices of the corresponding 2D spectra.}
\label{fig:exp_actual}
\end{figure}

Having validated the algorithm with simulated spectra, we then apply the network to analyze experimental 2D spectra. As an example, two experimental spectra obtained from a potassium (K) atomic vapor and a GaAs quantum well, as shown in Fig. \ref{fig:exp_actual}(a) and (c) respectively, were considered. The two spectra were analyzed by the trained neural network which extracted the linewidth values $\gamma=0.038$ THz and $\sigma=0.008$ THz for the K vapor, and $\gamma=0.045$ THz and $\sigma=0.103$ THz for the quantum well. These extracted linewidths were used in Eq. (\ref{eq1}) to generate two simulated spectra as shown in Fig. \ref{fig:exp_actual}(b) for the K vapor and Fig. \ref{fig:exp_actual}(d) for the quantum well. To evaluate the accuracy of the extracted linewidths, the experimental and simulated spectra are compared by using their corresponding diagonal and cross-diagonal slices, which are plotted in Fig. \ref{fig:exp_actual}(e) and (f), respectively, for the K vapor and in Fig. \ref{fig:exp_actual}(g) and (h), respectively, for the quantum well. The black dots are the slices from the experimental spectra and the red lines are the slices from the simulated spectra using the extracted linewidth values. The slices from simulation agree well with the experimental slices, demonstrating the accuracy of the extracted linewidths by the trained neural network. 

In summary, we implemented a machine learning algorithm based on a convolutional neural network to analyze 2D spectra for linewidth information. After training with a set of simulated 2D spectra with known linewidth values, the neural network can analyze a different set of simulated 2D spectra and two experimental 2D spectra to extract both homogeneous and inhomogeneous linewidths. The outputs are reliable and accurate. This approach can be generalized to analyze 2D spectra with more sophisticated spectral features to extract more parameters. The machine learning algorithm can be a good alternative in the cases that are usually difficult for conventional fitting.

\textbf{Funding.}
H.L. acknowledges support from the National Science Foundation (NSF) (PHY-1707364). Sandia National Laboratories is a multimission laboratory managed and operated by National Technology \& Engineering Solutions of Sandia, LLC, a wholly owned subsidiary of Honeywell International Inc., for the U.S. Department of Energy’s National Nuclear Security Administration under contract DE-NA0003525. This paper describes objective technical results and analysis. Any subjective views or opinions that might be expressed in the paper do not necessarily represent the views of the U.S. Department of Energy or the United States Government.

\bibliography{MLLinewidth}

\begin{thebibliography}{45}
\expandafter\ifx\csname natexlab\endcsname\relax\def\natexlab#1{#1}\fi
\expandafter\ifx\csname bibnamefont\endcsname\relax
  \def\bibnamefont#1{#1}\fi
\expandafter\ifx\csname bibfnamefont\endcsname\relax
  \def\bibfnamefont#1{#1}\fi
\expandafter\ifx\csname citenamefont\endcsname\relax
  \def\citenamefont#1{#1}\fi
\expandafter\ifx\csname url\endcsname\relax
  \def\url#1{\texttt{#1}}\fi
\expandafter\ifx\csname urlprefix\endcsname\relax\def\urlprefix{URL }\fi
\providecommand{\bibinfo}[2]{#2}
\providecommand{\eprint}[2][]{\url{#2}}

\bibitem[{\citenamefont{Cundiff and Mukamel}(2013)}]{Cundiff2013}
\bibinfo{author}{\bibfnamefont{S.~T.} \bibnamefont{Cundiff}} \bibnamefont{and}
  \bibinfo{author}{\bibfnamefont{S.}~\bibnamefont{Mukamel}},
  \bibinfo{journal}{Phys. Today} \textbf{\bibinfo{volume}{66}},
  \bibinfo{pages}{44} (\bibinfo{year}{2013}), ISSN \bibinfo{issn}{00319228}.

\bibitem[{\citenamefont{Li and Cundiff}(2017)}]{Li2017}
\bibinfo{author}{\bibfnamefont{H.}~\bibnamefont{Li}} \bibnamefont{and}
  \bibinfo{author}{\bibfnamefont{S.~T.} \bibnamefont{Cundiff}}, in
  \emph{\bibinfo{booktitle}{Advances In Atomic, Molecular, and Optical
  Physics}} (\bibinfo{publisher}{Elsevier}, \bibinfo{year}{2017}), pp.
  \bibinfo{pages}{1--48}.

\bibitem[{\citenamefont{Tian et~al.}(2003)\citenamefont{Tian, Keusters, Suzaki,
  and Warren}}]{Tian2003}
\bibinfo{author}{\bibfnamefont{P.~F.} \bibnamefont{Tian}},
  \bibinfo{author}{\bibfnamefont{D.}~\bibnamefont{Keusters}},
  \bibinfo{author}{\bibfnamefont{Y.}~\bibnamefont{Suzaki}}, \bibnamefont{and}
  \bibinfo{author}{\bibfnamefont{W.~S.} \bibnamefont{Warren}},
  \bibinfo{journal}{Science} \textbf{\bibinfo{volume}{300}},
  \bibinfo{pages}{1553} (\bibinfo{year}{2003}), ISSN \bibinfo{issn}{1095-9203}.

\bibitem[{\citenamefont{Tekavec et~al.}(2007)\citenamefont{Tekavec, Lott, and
  Marcus}}]{Tekavec2007}
\bibinfo{author}{\bibfnamefont{P.~F.} \bibnamefont{Tekavec}},
  \bibinfo{author}{\bibfnamefont{G.~A.} \bibnamefont{Lott}}, \bibnamefont{and}
  \bibinfo{author}{\bibfnamefont{A.~H.} \bibnamefont{Marcus}},
  \bibinfo{journal}{J. Chem. Phys.} \textbf{\bibinfo{volume}{127}},
  \bibinfo{pages}{214307} (\bibinfo{year}{2007}).

\bibitem[{\citenamefont{Dai et~al.}(2012)\citenamefont{Dai, Richter, Li,
  Bristow, Falvo, Mukamel, and Cundiff}}]{Dai2012}
\bibinfo{author}{\bibfnamefont{X.}~\bibnamefont{Dai}},
  \bibinfo{author}{\bibfnamefont{M.}~\bibnamefont{Richter}},
  \bibinfo{author}{\bibfnamefont{H.}~\bibnamefont{Li}},
  \bibinfo{author}{\bibfnamefont{A.~D.} \bibnamefont{Bristow}},
  \bibinfo{author}{\bibfnamefont{C.}~\bibnamefont{Falvo}},
  \bibinfo{author}{\bibfnamefont{S.}~\bibnamefont{Mukamel}}, \bibnamefont{and}
  \bibinfo{author}{\bibfnamefont{S.~T.} \bibnamefont{Cundiff}},
  \bibinfo{journal}{Phys. Rev. Lett.} \textbf{\bibinfo{volume}{108}},
  \bibinfo{pages}{193201} (\bibinfo{year}{2012}), ISSN
  \bibinfo{issn}{0031-9007}.

\bibitem[{\citenamefont{Li et~al.}(2013)\citenamefont{Li, Bristow, Siemens,
  Moody, and Cundiff}}]{Li2013}
\bibinfo{author}{\bibfnamefont{H.}~\bibnamefont{Li}},
  \bibinfo{author}{\bibfnamefont{A.~D.} \bibnamefont{Bristow}},
  \bibinfo{author}{\bibfnamefont{M.~E.} \bibnamefont{Siemens}},
  \bibinfo{author}{\bibfnamefont{G.}~\bibnamefont{Moody}}, \bibnamefont{and}
  \bibinfo{author}{\bibfnamefont{S.~T.} \bibnamefont{Cundiff}},
  \bibinfo{journal}{Nat. Commun.} \textbf{\bibinfo{volume}{4}},
  \bibinfo{pages}{1390} (\bibinfo{year}{2013}), ISSN \bibinfo{issn}{2041-1723}.

\bibitem[{\citenamefont{Gao et~al.}(2016)\citenamefont{Gao, Cundiff, and
  Li}}]{Gao:16}
\bibinfo{author}{\bibfnamefont{F.}~\bibnamefont{Gao}},
  \bibinfo{author}{\bibfnamefont{S.~T.} \bibnamefont{Cundiff}},
  \bibnamefont{and} \bibinfo{author}{\bibfnamefont{H.}~\bibnamefont{Li}},
  \bibinfo{journal}{Opt. Lett.} \textbf{\bibinfo{volume}{41}},
  \bibinfo{pages}{2954} (\bibinfo{year}{2016}).

\bibitem[{\citenamefont{Lomsadze and Cundiff}(2018)}]{PhysRevLett.120.233401}
\bibinfo{author}{\bibfnamefont{B.}~\bibnamefont{Lomsadze}} \bibnamefont{and}
  \bibinfo{author}{\bibfnamefont{S.~T.} \bibnamefont{Cundiff}},
  \bibinfo{journal}{Phys. Rev. Lett.} \textbf{\bibinfo{volume}{120}},
  \bibinfo{pages}{233401} (\bibinfo{year}{2018}).

\bibitem[{\citenamefont{Yu et~al.}(2019{\natexlab{a}})\citenamefont{Yu, Titze,
  Zhu, Liu, and Li}}]{Yu2019}
\bibinfo{author}{\bibfnamefont{S.}~\bibnamefont{Yu}},
  \bibinfo{author}{\bibfnamefont{M.}~\bibnamefont{Titze}},
  \bibinfo{author}{\bibfnamefont{Y.}~\bibnamefont{Zhu}},
  \bibinfo{author}{\bibfnamefont{X.}~\bibnamefont{Liu}}, \bibnamefont{and}
  \bibinfo{author}{\bibfnamefont{H.}~\bibnamefont{Li}}, \bibinfo{journal}{Opt.
  Lett.} \textbf{\bibinfo{volume}{44}}, \bibinfo{pages}{2795}
  (\bibinfo{year}{2019}{\natexlab{a}}), ISSN \bibinfo{issn}{0146-9592}.

\bibitem[{\citenamefont{Yu et~al.}(2019{\natexlab{b}})\citenamefont{Yu, Titze,
  Zhu, Liu, and Li}}]{Yu2018}
\bibinfo{author}{\bibfnamefont{S.}~\bibnamefont{Yu}},
  \bibinfo{author}{\bibfnamefont{M.}~\bibnamefont{Titze}},
  \bibinfo{author}{\bibfnamefont{Y.}~\bibnamefont{Zhu}},
  \bibinfo{author}{\bibfnamefont{X.}~\bibnamefont{Liu}}, \bibnamefont{and}
  \bibinfo{author}{\bibfnamefont{H.}~\bibnamefont{Li}}, \bibinfo{journal}{Opt.
  Express} \textbf{\bibinfo{volume}{27}}, \bibinfo{pages}{28891}
  (\bibinfo{year}{2019}{\natexlab{b}}).

\bibitem[{\citenamefont{Li et~al.}(2006)\citenamefont{Li, Zhang, Borca, and
  Cundiff}}]{Li2006a}
\bibinfo{author}{\bibfnamefont{X.}~\bibnamefont{Li}},
  \bibinfo{author}{\bibfnamefont{T.}~\bibnamefont{Zhang}},
  \bibinfo{author}{\bibfnamefont{C.}~\bibnamefont{Borca}}, \bibnamefont{and}
  \bibinfo{author}{\bibfnamefont{S.}~\bibnamefont{Cundiff}},
  \bibinfo{journal}{Phys. Rev. Lett.} \textbf{\bibinfo{volume}{96}},
  \bibinfo{pages}{057406} (\bibinfo{year}{2006}), ISSN
  \bibinfo{issn}{0031-9007}.

\bibitem[{\citenamefont{Stone et~al.}(2009)\citenamefont{Stone, Gundogdu,
  Turner, Li, Cundiff, and Nelson}}]{Stone2009a}
\bibinfo{author}{\bibfnamefont{K.~W.} \bibnamefont{Stone}},
  \bibinfo{author}{\bibfnamefont{K.}~\bibnamefont{Gundogdu}},
  \bibinfo{author}{\bibfnamefont{D.~B.} \bibnamefont{Turner}},
  \bibinfo{author}{\bibfnamefont{X.}~\bibnamefont{Li}},
  \bibinfo{author}{\bibfnamefont{S.~T.} \bibnamefont{Cundiff}},
  \bibnamefont{and} \bibinfo{author}{\bibfnamefont{K.~a.}
  \bibnamefont{Nelson}}, \bibinfo{journal}{Science}
  \textbf{\bibinfo{volume}{324}}, \bibinfo{pages}{1169} (\bibinfo{year}{2009}),
  ISSN \bibinfo{issn}{0036-8075}.

\bibitem[{\citenamefont{Cundiff et~al.}(2012)\citenamefont{Cundiff, Bristow,
  Siemens, Li, Moody, Karaiskaj, Dai, and Zhang}}]{Cundiff2012}
\bibinfo{author}{\bibfnamefont{S.~T.} \bibnamefont{Cundiff}},
  \bibinfo{author}{\bibfnamefont{A.~D.} \bibnamefont{Bristow}},
  \bibinfo{author}{\bibfnamefont{M.}~\bibnamefont{Siemens}},
  \bibinfo{author}{\bibfnamefont{H.}~\bibnamefont{Li}},
  \bibinfo{author}{\bibfnamefont{G.}~\bibnamefont{Moody}},
  \bibinfo{author}{\bibfnamefont{D.}~\bibnamefont{Karaiskaj}},
  \bibinfo{author}{\bibfnamefont{X.}~\bibnamefont{Dai}}, \bibnamefont{and}
  \bibinfo{author}{\bibfnamefont{T.}~\bibnamefont{Zhang}},
  \bibinfo{journal}{IEEE J. Sel. Top. Quantum Electron.}
  \textbf{\bibinfo{volume}{18}}, \bibinfo{pages}{318} (\bibinfo{year}{2012}),
  ISSN \bibinfo{issn}{1077260X}.

\bibitem[{\citenamefont{Turner et~al.}(2012)\citenamefont{Turner, Wen, Arias,
  Nelson, Li, Moody, Siemens, and Cundiff}}]{Turner2012}
\bibinfo{author}{\bibfnamefont{D.}~\bibnamefont{Turner}},
  \bibinfo{author}{\bibfnamefont{P.}~\bibnamefont{Wen}},
  \bibinfo{author}{\bibfnamefont{D.}~\bibnamefont{Arias}},
  \bibinfo{author}{\bibfnamefont{K.}~\bibnamefont{Nelson}},
  \bibinfo{author}{\bibfnamefont{H.}~\bibnamefont{Li}},
  \bibinfo{author}{\bibfnamefont{G.}~\bibnamefont{Moody}},
  \bibinfo{author}{\bibfnamefont{M.}~\bibnamefont{Siemens}}, \bibnamefont{and}
  \bibinfo{author}{\bibfnamefont{S.}~\bibnamefont{Cundiff}},
  \bibinfo{journal}{Phys. Rev. B} \textbf{\bibinfo{volume}{85}},
  \bibinfo{pages}{201303} (\bibinfo{year}{2012}), ISSN
  \bibinfo{issn}{1098-0121}.

\bibitem[{\citenamefont{Singh et~al.}(2013)\citenamefont{Singh, Autry, Nardin,
  Moody, Li, Pierz, Bieler, and Cundiff}}]{Singh2013}
\bibinfo{author}{\bibfnamefont{R.}~\bibnamefont{Singh}},
  \bibinfo{author}{\bibfnamefont{T.~M.} \bibnamefont{Autry}},
  \bibinfo{author}{\bibfnamefont{G.}~\bibnamefont{Nardin}},
  \bibinfo{author}{\bibfnamefont{G.}~\bibnamefont{Moody}},
  \bibinfo{author}{\bibfnamefont{H.}~\bibnamefont{Li}},
  \bibinfo{author}{\bibfnamefont{K.}~\bibnamefont{Pierz}},
  \bibinfo{author}{\bibfnamefont{M.}~\bibnamefont{Bieler}}, \bibnamefont{and}
  \bibinfo{author}{\bibfnamefont{S.~T.} \bibnamefont{Cundiff}},
  \bibinfo{journal}{Phys. Rev. B} \textbf{\bibinfo{volume}{88}},
  \bibinfo{pages}{45304} (\bibinfo{year}{2013}).

\bibitem[{\citenamefont{Moody et~al.}(2014)\citenamefont{Moody, Akimov, Li,
  Singh, Yakovlev, Karczewski, Wiater, Wojtowicz, Bayer, and
  Cundiff}}]{PhysRevLett.112.097401}
\bibinfo{author}{\bibfnamefont{G.}~\bibnamefont{Moody}},
  \bibinfo{author}{\bibfnamefont{I.~A.} \bibnamefont{Akimov}},
  \bibinfo{author}{\bibfnamefont{H.}~\bibnamefont{Li}},
  \bibinfo{author}{\bibfnamefont{R.}~\bibnamefont{Singh}},
  \bibinfo{author}{\bibfnamefont{D.~R.} \bibnamefont{Yakovlev}},
  \bibinfo{author}{\bibfnamefont{G.}~\bibnamefont{Karczewski}},
  \bibinfo{author}{\bibfnamefont{M.}~\bibnamefont{Wiater}},
  \bibinfo{author}{\bibfnamefont{T.}~\bibnamefont{Wojtowicz}},
  \bibinfo{author}{\bibfnamefont{M.}~\bibnamefont{Bayer}}, \bibnamefont{and}
  \bibinfo{author}{\bibfnamefont{S.~T.} \bibnamefont{Cundiff}},
  \bibinfo{journal}{Phys. Rev. Lett.} \textbf{\bibinfo{volume}{112}},
  \bibinfo{pages}{97401} (\bibinfo{year}{2014}).

\bibitem[{\citenamefont{Nardin et~al.}(2014)\citenamefont{Nardin, Moody, Singh,
  Autry, Li, Morier-Genoud, and Cundiff}}]{Nardin2014}
\bibinfo{author}{\bibfnamefont{G.}~\bibnamefont{Nardin}},
  \bibinfo{author}{\bibfnamefont{G.}~\bibnamefont{Moody}},
  \bibinfo{author}{\bibfnamefont{R.}~\bibnamefont{Singh}},
  \bibinfo{author}{\bibfnamefont{T.~M.} \bibnamefont{Autry}},
  \bibinfo{author}{\bibfnamefont{H.}~\bibnamefont{Li}},
  \bibinfo{author}{\bibfnamefont{F.}~\bibnamefont{Morier-Genoud}},
  \bibnamefont{and} \bibinfo{author}{\bibfnamefont{S.~T.}
  \bibnamefont{Cundiff}}, \bibinfo{journal}{Phys. Rev. Lett.}
  \textbf{\bibinfo{volume}{112}}, \bibinfo{pages}{046402}
  (\bibinfo{year}{2014}), ISSN \bibinfo{issn}{0031-9007}, \eprint{1308.1689}.

\bibitem[{\citenamefont{Moody et~al.}(2013{\natexlab{a}})\citenamefont{Moody,
  Singh, Li, Akimov, Bayer, Reuter, Wieck, Bracker, Gammon, and
  Cundiff}}]{PhysRevB.87.041304}
\bibinfo{author}{\bibfnamefont{G.}~\bibnamefont{Moody}},
  \bibinfo{author}{\bibfnamefont{R.}~\bibnamefont{Singh}},
  \bibinfo{author}{\bibfnamefont{H.}~\bibnamefont{Li}},
  \bibinfo{author}{\bibfnamefont{I.~A.} \bibnamefont{Akimov}},
  \bibinfo{author}{\bibfnamefont{M.}~\bibnamefont{Bayer}},
  \bibinfo{author}{\bibfnamefont{D.}~\bibnamefont{Reuter}},
  \bibinfo{author}{\bibfnamefont{A.~D.} \bibnamefont{Wieck}},
  \bibinfo{author}{\bibfnamefont{A.~S.} \bibnamefont{Bracker}},
  \bibinfo{author}{\bibfnamefont{D.}~\bibnamefont{Gammon}}, \bibnamefont{and}
  \bibinfo{author}{\bibfnamefont{S.~T.} \bibnamefont{Cundiff}},
  \bibinfo{journal}{Phys. Rev. B} \textbf{\bibinfo{volume}{87}},
  \bibinfo{pages}{41304} (\bibinfo{year}{2013}{\natexlab{a}}).

\bibitem[{\citenamefont{Moody et~al.}(2013{\natexlab{b}})\citenamefont{Moody,
  Singh, Li, Akimov, Bayer, Reuter, Wieck, and Cundiff}}]{Moody2013}
\bibinfo{author}{\bibfnamefont{G.}~\bibnamefont{Moody}},
  \bibinfo{author}{\bibfnamefont{R.}~\bibnamefont{Singh}},
  \bibinfo{author}{\bibfnamefont{H.}~\bibnamefont{Li}},
  \bibinfo{author}{\bibfnamefont{I.~A.} \bibnamefont{Akimov}},
  \bibinfo{author}{\bibfnamefont{M.}~\bibnamefont{Bayer}},
  \bibinfo{author}{\bibfnamefont{D.}~\bibnamefont{Reuter}},
  \bibinfo{author}{\bibfnamefont{A.~D.} \bibnamefont{Wieck}}, \bibnamefont{and}
  \bibinfo{author}{\bibfnamefont{S.~T.} \bibnamefont{Cundiff}},
  \bibinfo{journal}{Phys. Rev. B} \textbf{\bibinfo{volume}{87}},
  \bibinfo{pages}{045313} (\bibinfo{year}{2013}{\natexlab{b}}), ISSN
  \bibinfo{issn}{10980121}, \eprint{arXiv:1210.8096v1}.

\bibitem[{\citenamefont{Moody et~al.}(2013{\natexlab{c}})\citenamefont{Moody,
  Singh, Li, Akimov, Bayer, Reuter, Wieck, and Cundiff}}]{Moody2013a}
\bibinfo{author}{\bibfnamefont{G.}~\bibnamefont{Moody}},
  \bibinfo{author}{\bibfnamefont{R.}~\bibnamefont{Singh}},
  \bibinfo{author}{\bibfnamefont{H.}~\bibnamefont{Li}},
  \bibinfo{author}{\bibfnamefont{I.}~\bibnamefont{Akimov}},
  \bibinfo{author}{\bibfnamefont{M.}~\bibnamefont{Bayer}},
  \bibinfo{author}{\bibfnamefont{D.}~\bibnamefont{Reuter}},
  \bibinfo{author}{\bibfnamefont{A.}~\bibnamefont{Wieck}}, \bibnamefont{and}
  \bibinfo{author}{\bibfnamefont{S.}~\bibnamefont{Cundiff}},
  \bibinfo{journal}{Solid State Commun.} \textbf{\bibinfo{volume}{163}},
  \bibinfo{pages}{65} (\bibinfo{year}{2013}{\natexlab{c}}), ISSN
  \bibinfo{issn}{00381098}, \eprint{arXiv:1212.6941v1}.

\bibitem[{\citenamefont{Moody et~al.}(2013{\natexlab{d}})\citenamefont{Moody,
  Singh, Li, Akimov, Bayer, Reuter, Wieck, Bracker, Gammon, and
  Cundiff}}]{Moody2013b}
\bibinfo{author}{\bibfnamefont{G.}~\bibnamefont{Moody}},
  \bibinfo{author}{\bibfnamefont{R.}~\bibnamefont{Singh}},
  \bibinfo{author}{\bibfnamefont{H.}~\bibnamefont{Li}},
  \bibinfo{author}{\bibfnamefont{I.~A.} \bibnamefont{Akimov}},
  \bibinfo{author}{\bibfnamefont{M.}~\bibnamefont{Bayer}},
  \bibinfo{author}{\bibfnamefont{D.}~\bibnamefont{Reuter}},
  \bibinfo{author}{\bibfnamefont{A.~D.} \bibnamefont{Wieck}},
  \bibinfo{author}{\bibfnamefont{A.~S.} \bibnamefont{Bracker}},
  \bibinfo{author}{\bibfnamefont{D.}~\bibnamefont{Gammon}}, \bibnamefont{and}
  \bibinfo{author}{\bibfnamefont{S.~T.} \bibnamefont{Cundiff}},
  \bibinfo{journal}{Phys. Status Solidi (b)} \textbf{\bibinfo{volume}{250}},
  \bibinfo{pages}{1753} (\bibinfo{year}{2013}{\natexlab{d}}), ISSN
  \bibinfo{issn}{03701972}.

\bibitem[{\citenamefont{Moody et~al.}(2015)\citenamefont{Moody, {Kavir Dass},
  Hao, Chen, Li, Singh, Tran, Clark, Xu, Bergh{\"{a}}user et~al.}}]{Moody2015}
\bibinfo{author}{\bibfnamefont{G.}~\bibnamefont{Moody}},
  \bibinfo{author}{\bibfnamefont{C.}~\bibnamefont{{Kavir Dass}}},
  \bibinfo{author}{\bibfnamefont{K.}~\bibnamefont{Hao}},
  \bibinfo{author}{\bibfnamefont{C.-H.} \bibnamefont{Chen}},
  \bibinfo{author}{\bibfnamefont{L.-J.} \bibnamefont{Li}},
  \bibinfo{author}{\bibfnamefont{A.}~\bibnamefont{Singh}},
  \bibinfo{author}{\bibfnamefont{K.}~\bibnamefont{Tran}},
  \bibinfo{author}{\bibfnamefont{G.}~\bibnamefont{Clark}},
  \bibinfo{author}{\bibfnamefont{X.}~\bibnamefont{Xu}},
  \bibinfo{author}{\bibfnamefont{G.}~\bibnamefont{Bergh{\"{a}}user}},
  \bibnamefont{et~al.}, \bibinfo{journal}{Nat. Commun.}
  \textbf{\bibinfo{volume}{6}}, \bibinfo{pages}{8315} (\bibinfo{year}{2015}),
  ISSN \bibinfo{issn}{2041-1723}.

\bibitem[{\citenamefont{Titze et~al.}(2018)\citenamefont{Titze, Li, Zhang,
  Ajayan, and Li}}]{Titze2018}
\bibinfo{author}{\bibfnamefont{M.}~\bibnamefont{Titze}},
  \bibinfo{author}{\bibfnamefont{B.}~\bibnamefont{Li}},
  \bibinfo{author}{\bibfnamefont{X.}~\bibnamefont{Zhang}},
  \bibinfo{author}{\bibfnamefont{P.~M.} \bibnamefont{Ajayan}},
  \bibnamefont{and} \bibinfo{author}{\bibfnamefont{H.}~\bibnamefont{Li}},
  \bibinfo{journal}{Phys. Rev. Materials} \textbf{\bibinfo{volume}{2}},
  \bibinfo{pages}{054001} (\bibinfo{year}{2018}), ISSN
  \bibinfo{issn}{2475-9953}.

\bibitem[{\citenamefont{Monahan et~al.}(2017)\citenamefont{Monahan, Guo, Lin,
  Dou, Yang, and Fleming}}]{Monahan2017}
\bibinfo{author}{\bibfnamefont{D.~M.} \bibnamefont{Monahan}},
  \bibinfo{author}{\bibfnamefont{L.}~\bibnamefont{Guo}},
  \bibinfo{author}{\bibfnamefont{J.}~\bibnamefont{Lin}},
  \bibinfo{author}{\bibfnamefont{L.}~\bibnamefont{Dou}},
  \bibinfo{author}{\bibfnamefont{P.}~\bibnamefont{Yang}}, \bibnamefont{and}
  \bibinfo{author}{\bibfnamefont{G.~R.} \bibnamefont{Fleming}},
  \bibinfo{journal}{J. Phys. Chem. Lett.} \textbf{\bibinfo{volume}{8}},
  \bibinfo{pages}{3211} (\bibinfo{year}{2017}).

\bibitem[{\citenamefont{Richter et~al.}(2017)\citenamefont{Richter, Branchi,
  Valduga~de Almeida~Camargo, Zhao, Friend, Cerullo, and
  Deschler}}]{Richter2017}
\bibinfo{author}{\bibfnamefont{J.~M.} \bibnamefont{Richter}},
  \bibinfo{author}{\bibfnamefont{F.}~\bibnamefont{Branchi}},
  \bibinfo{author}{\bibfnamefont{F.}~\bibnamefont{Valduga~de Almeida~Camargo}},
  \bibinfo{author}{\bibfnamefont{B.}~\bibnamefont{Zhao}},
  \bibinfo{author}{\bibfnamefont{R.~H.} \bibnamefont{Friend}},
  \bibinfo{author}{\bibfnamefont{G.}~\bibnamefont{Cerullo}}, \bibnamefont{and}
  \bibinfo{author}{\bibfnamefont{F.}~\bibnamefont{Deschler}},
  \bibinfo{journal}{Nat. Commun.} \textbf{\bibinfo{volume}{8}},
  \bibinfo{pages}{376} (\bibinfo{year}{2017}).

\bibitem[{\citenamefont{Jha et~al.}(2018)\citenamefont{Jha, Duan, Tiwari,
  Nayak, Snaith, Thorwart, and Miller}}]{Jha2018}
\bibinfo{author}{\bibfnamefont{A.}~\bibnamefont{Jha}},
  \bibinfo{author}{\bibfnamefont{H.-G.} \bibnamefont{Duan}},
  \bibinfo{author}{\bibfnamefont{V.}~\bibnamefont{Tiwari}},
  \bibinfo{author}{\bibfnamefont{P.~K.} \bibnamefont{Nayak}},
  \bibinfo{author}{\bibfnamefont{H.~J.} \bibnamefont{Snaith}},
  \bibinfo{author}{\bibfnamefont{M.}~\bibnamefont{Thorwart}}, \bibnamefont{and}
  \bibinfo{author}{\bibfnamefont{R.~J.~D.} \bibnamefont{Miller}},
  \bibinfo{journal}{ACS Photonics} \textbf{\bibinfo{volume}{5}},
  \bibinfo{pages}{852} (\bibinfo{year}{2018}).

\bibitem[{\citenamefont{Thouin et~al.}(2018)\citenamefont{Thouin, Neutzner,
  Cortecchia, Dragomir, Soci, Salim, Lam, Leonelli, Petrozza, Kandada
  et~al.}}]{Thouin2018}
\bibinfo{author}{\bibfnamefont{F.}~\bibnamefont{Thouin}},
  \bibinfo{author}{\bibfnamefont{S.}~\bibnamefont{Neutzner}},
  \bibinfo{author}{\bibfnamefont{D.}~\bibnamefont{Cortecchia}},
  \bibinfo{author}{\bibfnamefont{V.~A.} \bibnamefont{Dragomir}},
  \bibinfo{author}{\bibfnamefont{C.}~\bibnamefont{Soci}},
  \bibinfo{author}{\bibfnamefont{T.}~\bibnamefont{Salim}},
  \bibinfo{author}{\bibfnamefont{Y.~M.} \bibnamefont{Lam}},
  \bibinfo{author}{\bibfnamefont{R.}~\bibnamefont{Leonelli}},
  \bibinfo{author}{\bibfnamefont{A.}~\bibnamefont{Petrozza}},
  \bibinfo{author}{\bibfnamefont{A.~R.~S.} \bibnamefont{Kandada}},
  \bibnamefont{et~al.}, \bibinfo{journal}{Phys. Rev. Materials}
  \textbf{\bibinfo{volume}{2}} (\bibinfo{year}{2018}).

\bibitem[{\citenamefont{Nishida et~al.}(2018)\citenamefont{Nishida, Breen,
  Lindquist, Umeyama, Karunadasa, and Fayer}}]{Nishida2018}
\bibinfo{author}{\bibfnamefont{J.}~\bibnamefont{Nishida}},
  \bibinfo{author}{\bibfnamefont{J.~P.} \bibnamefont{Breen}},
  \bibinfo{author}{\bibfnamefont{K.~P.} \bibnamefont{Lindquist}},
  \bibinfo{author}{\bibfnamefont{D.}~\bibnamefont{Umeyama}},
  \bibinfo{author}{\bibfnamefont{H.~I.} \bibnamefont{Karunadasa}},
  \bibnamefont{and} \bibinfo{author}{\bibfnamefont{M.~D.} \bibnamefont{Fayer}},
  \bibinfo{journal}{J. Am. Chem. Soc.} \textbf{\bibinfo{volume}{140}},
  \bibinfo{pages}{9882} (\bibinfo{year}{2018}).

\bibitem[{\citenamefont{Titze et~al.}(2019)\citenamefont{Titze, Fei, Munoz,
  Wang, Wang, and Li}}]{Titze2019}
\bibinfo{author}{\bibfnamefont{M.}~\bibnamefont{Titze}},
  \bibinfo{author}{\bibfnamefont{C.}~\bibnamefont{Fei}},
  \bibinfo{author}{\bibfnamefont{M.}~\bibnamefont{Munoz}},
  \bibinfo{author}{\bibfnamefont{X.}~\bibnamefont{Wang}},
  \bibinfo{author}{\bibfnamefont{H.}~\bibnamefont{Wang}}, \bibnamefont{and}
  \bibinfo{author}{\bibfnamefont{H.}~\bibnamefont{Li}}, \bibinfo{journal}{J.
  Phys. Chem. Lett.} \textbf{\bibinfo{volume}{10}}, \bibinfo{pages}{4625}
  (\bibinfo{year}{2019}).

\bibitem[{\citenamefont{Brixner et~al.}(2005)\citenamefont{Brixner, Stenger,
  Vaswani, Cho, Blankenship, and Fleming}}]{Brixner2005}
\bibinfo{author}{\bibfnamefont{T.}~\bibnamefont{Brixner}},
  \bibinfo{author}{\bibfnamefont{J.}~\bibnamefont{Stenger}},
  \bibinfo{author}{\bibfnamefont{H.~M.} \bibnamefont{Vaswani}},
  \bibinfo{author}{\bibfnamefont{M.}~\bibnamefont{Cho}},
  \bibinfo{author}{\bibfnamefont{R.~E.} \bibnamefont{Blankenship}},
  \bibnamefont{and} \bibinfo{author}{\bibfnamefont{G.~R.}
  \bibnamefont{Fleming}}, \bibinfo{journal}{Nature}
  \textbf{\bibinfo{volume}{434}}, \bibinfo{pages}{625} (\bibinfo{year}{2005}),
  ISSN \bibinfo{issn}{0028-0836}.

\bibitem[{\citenamefont{Engel et~al.}(2007)\citenamefont{Engel, Calhoun, Read,
  Ahn, Man{\v{c}}al, Cheng, Blankenship, and Fleming}}]{Engel2007}
\bibinfo{author}{\bibfnamefont{G.~S.} \bibnamefont{Engel}},
  \bibinfo{author}{\bibfnamefont{T.~R.} \bibnamefont{Calhoun}},
  \bibinfo{author}{\bibfnamefont{E.~L.} \bibnamefont{Read}},
  \bibinfo{author}{\bibfnamefont{T.-K.} \bibnamefont{Ahn}},
  \bibinfo{author}{\bibfnamefont{T.}~\bibnamefont{Man{\v{c}}al}},
  \bibinfo{author}{\bibfnamefont{Y.-C.} \bibnamefont{Cheng}},
  \bibinfo{author}{\bibfnamefont{R.~E.} \bibnamefont{Blankenship}},
  \bibnamefont{and} \bibinfo{author}{\bibfnamefont{G.~R.}
  \bibnamefont{Fleming}}, \bibinfo{journal}{Nature}
  \textbf{\bibinfo{volume}{446}}, \bibinfo{pages}{782} (\bibinfo{year}{2007}),
  ISSN \bibinfo{issn}{0028-0836}.

\bibitem[{\citenamefont{Collini et~al.}(2010)\citenamefont{Collini, Wong, Wilk,
  Curmi, Brumer, and Scholes}}]{Collini2010}
\bibinfo{author}{\bibfnamefont{E.}~\bibnamefont{Collini}},
  \bibinfo{author}{\bibfnamefont{C.~Y.} \bibnamefont{Wong}},
  \bibinfo{author}{\bibfnamefont{K.~E.} \bibnamefont{Wilk}},
  \bibinfo{author}{\bibfnamefont{P.~M.~G.} \bibnamefont{Curmi}},
  \bibinfo{author}{\bibfnamefont{P.}~\bibnamefont{Brumer}}, \bibnamefont{and}
  \bibinfo{author}{\bibfnamefont{G.~D.} \bibnamefont{Scholes}},
  \bibinfo{journal}{Nature} \textbf{\bibinfo{volume}{463}},
  \bibinfo{pages}{644} (\bibinfo{year}{2010}), ISSN \bibinfo{issn}{1476-4687}.

\bibitem[{\citenamefont{Siemens et~al.}(2010)\citenamefont{Siemens, Moody, Li,
  Bristow, and Cundiff}}]{Siemens2010}
\bibinfo{author}{\bibfnamefont{M.~E.} \bibnamefont{Siemens}},
  \bibinfo{author}{\bibfnamefont{G.}~\bibnamefont{Moody}},
  \bibinfo{author}{\bibfnamefont{H.}~\bibnamefont{Li}},
  \bibinfo{author}{\bibfnamefont{A.~D.} \bibnamefont{Bristow}},
  \bibnamefont{and} \bibinfo{author}{\bibfnamefont{S.~T.}
  \bibnamefont{Cundiff}}, \bibinfo{journal}{Opt. Express}
  \textbf{\bibinfo{volume}{18}}, \bibinfo{pages}{17699} (\bibinfo{year}{2010}).

\bibitem[{\citenamefont{Bell et~al.}(2015)\citenamefont{Bell, Conrad, and
  Siemens}}]{Bell2015}
\bibinfo{author}{\bibfnamefont{J.~D.} \bibnamefont{Bell}},
  \bibinfo{author}{\bibfnamefont{R.}~\bibnamefont{Conrad}}, \bibnamefont{and}
  \bibinfo{author}{\bibfnamefont{M.~E.} \bibnamefont{Siemens}},
  \bibinfo{journal}{Opt. Lett.} \textbf{\bibinfo{volume}{40}},
  \bibinfo{pages}{1157} (\bibinfo{year}{2015}).

\bibitem[{\citenamefont{Carleo and Troyer}(2017)}]{Carleo2017}
\bibinfo{author}{\bibfnamefont{G.}~\bibnamefont{Carleo}} \bibnamefont{and}
  \bibinfo{author}{\bibfnamefont{M.}~\bibnamefont{Troyer}},
  \bibinfo{journal}{Science} \textbf{\bibinfo{volume}{355}},
  \bibinfo{pages}{602} (\bibinfo{year}{2017}).

\bibitem[{\citenamefont{Torlai et~al.}(2018)\citenamefont{Torlai, Mazzola,
  Carrasquilla, Troyer, Melko, and Carleo}}]{Torlai2018}
\bibinfo{author}{\bibfnamefont{G.}~\bibnamefont{Torlai}},
  \bibinfo{author}{\bibfnamefont{G.}~\bibnamefont{Mazzola}},
  \bibinfo{author}{\bibfnamefont{J.}~\bibnamefont{Carrasquilla}},
  \bibinfo{author}{\bibfnamefont{M.}~\bibnamefont{Troyer}},
  \bibinfo{author}{\bibfnamefont{R.}~\bibnamefont{Melko}}, \bibnamefont{and}
  \bibinfo{author}{\bibfnamefont{G.}~\bibnamefont{Carleo}},
  \bibinfo{journal}{Nat. Phys.} \textbf{\bibinfo{volume}{14}},
  \bibinfo{pages}{447} (\bibinfo{year}{2018}).

\bibitem[{\citenamefont{Xue et~al.}(2016)\citenamefont{Xue, Balachandran,
  Hogden, Theiler, Xue, and Lookman}}]{Xue2016}
\bibinfo{author}{\bibfnamefont{D.}~\bibnamefont{Xue}},
  \bibinfo{author}{\bibfnamefont{P.~V.} \bibnamefont{Balachandran}},
  \bibinfo{author}{\bibfnamefont{J.}~\bibnamefont{Hogden}},
  \bibinfo{author}{\bibfnamefont{J.}~\bibnamefont{Theiler}},
  \bibinfo{author}{\bibfnamefont{D.}~\bibnamefont{Xue}}, \bibnamefont{and}
  \bibinfo{author}{\bibfnamefont{T.}~\bibnamefont{Lookman}},
  \bibinfo{journal}{Nat. Commun.} \textbf{\bibinfo{volume}{7}},
  \bibinfo{pages}{11241} (\bibinfo{year}{2016}).

\bibitem[{\citenamefont{Ghosh et~al.}(2019)\citenamefont{Ghosh, Stuke,
  Todorovi{\'{c}}, J{\o}rgensen, Schmidt, Vehtari, and Rinke}}]{Ghosh2019}
\bibinfo{author}{\bibfnamefont{K.}~\bibnamefont{Ghosh}},
  \bibinfo{author}{\bibfnamefont{A.}~\bibnamefont{Stuke}},
  \bibinfo{author}{\bibfnamefont{M.}~\bibnamefont{Todorovi{\'{c}}}},
  \bibinfo{author}{\bibfnamefont{P.~B.} \bibnamefont{J{\o}rgensen}},
  \bibinfo{author}{\bibfnamefont{M.~N.} \bibnamefont{Schmidt}},
  \bibinfo{author}{\bibfnamefont{A.}~\bibnamefont{Vehtari}}, \bibnamefont{and}
  \bibinfo{author}{\bibfnamefont{P.}~\bibnamefont{Rinke}},
  \bibinfo{journal}{Adv. Sci.} \textbf{\bibinfo{volume}{6}},
  \bibinfo{pages}{1801367} (\bibinfo{year}{2019}).

\bibitem[{\citenamefont{Yajima and Taira}(1979)}]{Yajima1979}
\bibinfo{author}{\bibfnamefont{T.}~\bibnamefont{Yajima}} \bibnamefont{and}
  \bibinfo{author}{\bibfnamefont{Y.}~\bibnamefont{Taira}}, \bibinfo{journal}{J.
  Phys. Soc. Jpn.} \textbf{\bibinfo{volume}{47}}, \bibinfo{pages}{1620}
  (\bibinfo{year}{1979}).

\bibitem[{\citenamefont{Goodfellow et~al.}(2016)\citenamefont{Goodfellow,
  Bengio, and Courville}}]{Goodfellow-et-al-2016}
\bibinfo{author}{\bibfnamefont{I.}~\bibnamefont{Goodfellow}},
  \bibinfo{author}{\bibfnamefont{Y.}~\bibnamefont{Bengio}}, \bibnamefont{and}
  \bibinfo{author}{\bibfnamefont{A.}~\bibnamefont{Courville}},
  \emph{\bibinfo{title}{Deep Learning}} (\bibinfo{publisher}{MIT Press},
  \bibinfo{year}{2016}), \bibinfo{note}{\url{http://www.deeplearningbook.org}}.

\bibitem[{\citenamefont{LeCun et~al.}(2015)\citenamefont{LeCun, Bengio, and
  Hinton}}]{lecun2015deep}
\bibinfo{author}{\bibfnamefont{Y.}~\bibnamefont{LeCun}},
  \bibinfo{author}{\bibfnamefont{Y.}~\bibnamefont{Bengio}}, \bibnamefont{and}
  \bibinfo{author}{\bibfnamefont{G.}~\bibnamefont{Hinton}},
  \bibinfo{journal}{Nature} \textbf{\bibinfo{volume}{521}},
  \bibinfo{pages}{436} (\bibinfo{year}{2015}).

\bibitem[{\citenamefont{Rodr{\'\i}guez and
  Kramer}(2019)}]{rodriguez2019machine}
\bibinfo{author}{\bibfnamefont{M.}~\bibnamefont{Rodr{\'\i}guez}}
  \bibnamefont{and} \bibinfo{author}{\bibfnamefont{T.}~\bibnamefont{Kramer}},
  \bibinfo{journal}{Chemical Physics} \textbf{\bibinfo{volume}{520}},
  \bibinfo{pages}{52} (\bibinfo{year}{2019}).

\bibitem[{\citenamefont{{Bhardwaj} et~al.}(2019)\citenamefont{{Bhardwaj},
  {Gohel}, and {Namuduri}}}]{8788580}
\bibinfo{author}{\bibfnamefont{S.}~\bibnamefont{{Bhardwaj}}},
  \bibinfo{author}{\bibfnamefont{H.}~\bibnamefont{{Gohel}}}, \bibnamefont{and}
  \bibinfo{author}{\bibfnamefont{S.}~\bibnamefont{{Namuduri}}},
  \bibinfo{journal}{IEEE Antennas and Wireless Propagation Letters}
  \textbf{\bibinfo{volume}{18}}, \bibinfo{pages}{1} (\bibinfo{year}{2019}).

\bibitem[{\citenamefont{LeCun et~al.}(1989)\citenamefont{LeCun, Boser, Denker,
  Henderson, Howard, Hubbard, and Jackel}}]{lecun1989backpropagation}
\bibinfo{author}{\bibfnamefont{Y.}~\bibnamefont{LeCun}},
  \bibinfo{author}{\bibfnamefont{B.}~\bibnamefont{Boser}},
  \bibinfo{author}{\bibfnamefont{J.~S.} \bibnamefont{Denker}},
  \bibinfo{author}{\bibfnamefont{D.}~\bibnamefont{Henderson}},
  \bibinfo{author}{\bibfnamefont{R.~E.} \bibnamefont{Howard}},
  \bibinfo{author}{\bibfnamefont{W.}~\bibnamefont{Hubbard}}, \bibnamefont{and}
  \bibinfo{author}{\bibfnamefont{L.~D.} \bibnamefont{Jackel}},
  \bibinfo{journal}{Neural Comput.} \textbf{\bibinfo{volume}{1}},
  \bibinfo{pages}{541} (\bibinfo{year}{1989}).

\bibitem[{\citenamefont{Bottou}(2010)}]{bottou2010large}
\bibinfo{author}{\bibfnamefont{L.}~\bibnamefont{Bottou}}, in
  \emph{\bibinfo{booktitle}{Proceedings of COMPSTAT'2010}}
  (\bibinfo{publisher}{Springer}, \bibinfo{year}{2010}), pp.
  \bibinfo{pages}{177--186}.

\end{thebibliography}

\end{document}